\documentstyle[12pt]{article}

\global\arraycolsep=1pt
\begin{document}
\date{TIT/HEP-247/COSMO-42 \\  August, 1994}
\title{Gravitational Instantons and Moduli Spaces   in \\
 Topological  2-form Gravity }
\author{
 {\bf Mitsuko Abe,}
\ \ {\bf  Akika Nakamichi ${}^{\dagger,} $}
\thanks{On leave of absence from Tokyo Institute of Technology
after April 1994.} \  and
\ \  \
{\bf  Tatsuya Ueno${}^{\ddagger,}{}^\ast$ }
 \\
{\it Department of Physics, Tokyo Institute of Technology} \\
{\it Oh-okayama, Meguro-ku, Tokyo 152, Japan } \\ \
{\it ${}^\dagger$ Uji Research Center,} \\
{\it Yukawa Institute for Theoretical Physics, Kyoto Uiversity, }\\
{\it   Uji 611, Japan}\\ \
{\it ${}^\ddagger$ Department of  Physics, Kyung Hee University}\\
{\it Tongdaemun-gu Seoul 130-701, Korea}
}
\maketitle
\abstract{
\par  A topological version of four-dimensional
(Euclidean) Einstein gravity which we propose
regards  anti-self-dual 2-forms and   an anti-self-dual part of
  the frame  connections  as fundamental fields.
 The theory describes the moduli spaces of conformally self-dual Einstein
manifolds for a cosmological constant $\Lambda  \not =0$ case
and Einstein-K\"ahlerian manifold with the  vanishing
real first Chern class  for $\Lambda  =0$.
 In the  $\Lambda  \not =0$ case,  we evaluate the index of
the elliptic complex associated with the moduli space and
calculate the partition function.
We also clarify the moduli space and its dimension for $\Lambda  =0$
which  are  related to the Plebansky's heavenly equations.}
\thispagestyle{empty}
\newpage
\bf{I.~~ INTRODUCTION }
\par
Recently a number of noteworthy connections have been revealed
between a class of field theories called topological quantum  field theories
on  one hand, and the mathematical advances in the topology and geometry of low
dimensional manifolds on the other.
The study of these relations has been introduced by
Schwarz\cite{schwarz} and Witten\cite{witten1}.
Topological quantum field theories are constructed
by fields, symmetries and equations.
\par
One concept that lies in the topological quantum
field theory is the realization of the moduli spaces.
The moduli space  is  defined as the  equivalent set of the solutions
 of the fields  for  the  equations associated to the symmetries of topological
 quantum theories.
These theories  can be described  by  the moduli spaces and
 are characterized by their  topological  and geometrical  invariants
which depend only on moduli parameters.
There may be various topological quantum field
 theories  which describe
the same moduli space.  The prime interest of these theories  is
these invariants, which are computable by standard techniques in quantum
field theories.
\par
Some gravitational versions of   topological quantum field theories are
also given by Witten\cite{witten2}{}$^,$\cite{witten3}.
The two-dimensional gravity models are of importance
and  promise  new  insight into the string  theory \cite{witten3}.
 He conjectured that  certain  series of critical
points in  the  matrix model approach
 (i.e. the dynamical simplicial decomposition of  Riemannian surfaces)
is equivalent to the two-dimensional topological gravity.
In fact, Kontsevich  used  the intersection theory \cite{kont}
to support the conjecture.
This result  is  important to  know the non-perturbative effect of the
string theory.
\par
Since the work of Witten, there have been several attempts to
construct  four-dimensional topological gravity theories  over
 different kinds of the gravitational moduli spaces \cite{perry}
${}^-$\cite{anselmi}.
For example, the moduli space of the  conformally self-dual gravitational
instantons was investigated in detail by Perry and Teo  \cite{perry}.
\par
In  previous papers \cite{lee}${}^,$\cite{abe} we   proposed a
four-dimensional  topological gravity model.
This model contains two types of  topological field theories
; Witten-type topological field theory in the
cosmological  constant $\Lambda \not= 0$ case
and Schwarz-type topological field theory in  the $\Lambda = 0$ case.
They are  obtained by modifying a chiral  formulation of Einstein gravity
developed by Capovilla et al.  \cite{capovilla}.
In these theories  three quaternionic
K\"ahler forms and the  anti-self-dual part of the frame  connections
 of the principal bundle   $P_{SO(4)}$  are used as  fundamental fields.
The advantages of using these fields  are   that   the
treatment  analogous to that of Yang-Mills field is possible and
 that the moduli  space can be  easily defined in terms of them  efficiently.
\par
In the  $\Lambda \not= 0$ case, the moduli space is the set of the equivalence
class of the fields defining  the Einstein conformally self-dual
Riemannian manifolds.
This moduli space is up to orientation, identical with the one
considered by Torre  \cite{torre}. In his paper the dimension of the moduli
space is found to be zero when the cosmological constant is positive, and the
result is true also in our case.
In the  $\Lambda = 0$ case, the moduli spaces are  those
 of   Einstein-K\"ahlerian manifolds with  vanishing
  real first Chern class.
\par
The purpose of our attempt is to investigate  the four-dimensional
gravitational instantons and derive
the topological invariants such as the partition function and
observables.  We  explore the relation between  the
simplicial decomposition of four-dimensional manifolds and
the four-dimensional topological gravity.
\par
We can regard the $\Lambda \not =0$ case as a simple example of
a gravitational analogue of the  Donaldson theory
 and expect that we can calculate  some
topological invariants such as  the partition function  in four dimensions.
\par
On the contrary, the $\Lambda=0$ case  is a BF-type
topological gravity model. The partition function of the
abelian  BF-theory  is represented by the Ray-Singer torsion \cite{blau}.
Thus it is interesting to confirm
whether  our partition function in the $\Lambda=0$  can be related to the
Ray-Singer torsion or not.
\par
Another aspect of the $\Lambda=0$ case  is that  it provides  the self-dual
equations of Riemannian curvature 2-form.
There have been discovered  various  kinds of non-compact gravitational
instantons (i.e. ALE  \cite{eguchi} or ALF \cite{hawking}  ) which satisfy
these equations.
In this paper we will treat  the compact manifolds  only. In the near future
we will extend our investigation to the non-compact case.
\par
The plan of the paper is as follows.
In section II, we  present a classical action, fields content and equations of
motion, and define the moduli spaces in our theory.
We explain each case  separately to avoid the confusion.
In section III,  we formulate  the BRST transformations of our model.
In section IV, we  mention the dimension of the moduli space and zero modes
which appear in the partition function in  the $\Lambda \not= 0$ case.
In section V, the gauge fixing conditions are introduced
and the partition function is derived in the $\Lambda \not= 0$ case.
In section VI, we explain  the dimension of the moduli space in the
$\Lambda = 0$ case.
  The section VII is devoted to discussion.
\newpage
\bf{II. TOPOLOGICAL 2-FORM GRAVITY}
\par
We  adopted the  following action suggested by Capovilla et al.
\cite{capovilla} and Horowitz  \cite{horowitz}
for our topological gravity model
on a  four-dimensional  manifold $M_4$ \cite{lee}$^,$\cite{abe}.
\begin{equation}
 S_{TG} = {1 \over \alpha} \int_{M_4} [\Sigma^k \wedge F_k
  - \frac{\Lambda}{24} \Sigma^k \wedge \Sigma_k]~~
{}~~~(k=1,~2,~3)  \, ,
\label{eq:one}
\end{equation}
 where  $\alpha$ is a dimensionless parameter and
 $\Lambda$   is a  cosmological constant
(as we will see later that it  will  appear in Einstein
equation $R_{ \mu \nu}= \Lambda g_{ \mu \nu}$
 with  $\mu, \nu=\{ 0, \cdots , 3 \}$ ).
\par
 We start with fundamental fields, a trio of su(2) valued  2-forms
$\Sigma^k $ and a
su(2) valued 1-form $\omega^k=\omega^k_\mu dx^\mu$.  $F^k$ denotes the
su(2) valued 2-form with  $F^k= F^k_{\mu\nu} dx^\mu
\wedge  dx^\nu \equiv
d\omega^k+ (\omega \times \omega)^k = d\omega^k + \epsilon^{ijk}
\omega^i \wedge \omega^j$
($\epsilon^{ijk}$ is  the structure constant of $SU(2)$).
 Varying the action with respect to each of fields $\Sigma^{k}$ and
$\omega^{k}$, we obtain the equations of motion:
{
\setcounter{enumi}{\value{equation}}
\addtocounter{enumi}{1}
\setcounter{equation}{0}
\renewcommand{\theequation}{\theenumi\alph{equation}}
\begin{eqnarray}
  &&\Lambda \not= 0~;~ F^k - {\Lambda \over 12} \,
\Sigma^k = 0,~~D{\Sigma^k} =  0
  \ ,
\label{eqn:ins1}
\\
  &&\Lambda=0~;~F^k=0,~~D\Sigma^k=0\ ,
                                        \label{eqn:ins2}
\end{eqnarray}
\setcounter{equation}{\value{enumi}}
}
where $D\Sigma^k=d{\Sigma^k} + 2 (\omega
\times \Sigma)^k$.
\par
In this paper, we take $\alpha \rightarrow 0$ limit in Eq. (\ref{eq:one})
to make the contribution from Eq.
(\ref{eqn:ins1}) or Eq. (\ref{eqn:ins2})  dominant in our theory.
We are interested in the moduli spaces which are defined by Eq.
(\ref{eqn:ins1}) or Eq. (\ref{eqn:ins2})
 and the gauge fixing conditions which we will explain later.
This treatment is  similar to that of the
large $k$-limit of the Chern-Simons theory \cite{witten4} .
\par
For $\Lambda \not= 0$, one of  equations of   motion
 $D\Sigma^k=0$  can be derived by
 $F^k - {\Lambda \over 12}\Sigma^k=0$
and Bianchi Identity $DF^k=0$.
Eliminating $\Sigma^k$ from the action
by using Eq. (\ref{eqn:ins1}) we obtain the effective action proportional
to  the second Chern number $\int F^k \wedge F_k $, which is the
classical action of the TYMT ( the Donaldson theory) for
the SU(2) gauge group  \cite{baulieu}. Thus the theory reduces to
a Witten-type topological gravity model  on-shell.
On the other hand, for the $\Lambda = 0$ case,
  the action describes a
Schwarz-type (BF-type) topological field theory
 \cite{blau}${}^,$\cite{horowitz}.
\\
\\
\par
We  suppose  the following conditions for the action.
\\
Postulate~ 1a~~ for  $M_4$   with $\Lambda \not= 0$~:
\par
$M_4 $  is a four-dimensional oriented Riemannian manifold.
\\
Postulate~ 1b~~ for  $M_4$   with  $\Lambda = 0$~:
\par
$M_4 $  is a four-dimensional oriented Riemannian manifold
             and has an almost complex structure
            with its real first Chern class $c_1(M_4)_R= 0$.
\\
Postulate ~2~~ for the field $\omega^k$~:
\par
We consider the principal bundle $P_{SO(4)}$  of  oriented orthonormal frames
 over $M_4$  with the structure group $SO(4)$.
This bundle is  associated   by the tangent bundle
with   a metric $\tilde g_{\mu \nu}
= \tilde e^a_\mu \tilde e^b_\nu \delta_{ab}$,
where   $a,b = \{ 0,   \dots 3 \} $.
 $\tilde e^a= \tilde  e^a_\mu dx^\mu$ is a vierbein (a section of
${\rm  End}(TM_4)=T^\ast M_4  \otimes TM_4$~~ with the  assumption of
det $ (\tilde  e) \not= 0 $~).
We  suppose that the field
 $\omega^k_\mu$ denotes  an anti-self-dual part of the frame connections
( a  connection of  $P_{SU(2)}$
 which comes from  $P_{SO(4)} \sim P_{SU(2)} +  P_{SU(2)}$).
 $\omega^k$   is related to anti-self-dual part of
 so(4) valued 1-form connection
${}^{(-)}\omega^{a b}$~via $\eta^k_{ab}$;
\begin{equation}{}^{(-)}\omega^{a b}_ \mu(\tilde e)= \eta_k^{a b}
\omega^k_\mu (\tilde e),
\label{eq:two}
\end{equation}
where  $\eta^k_{ab}$ is an anti-self-dual constant  called the t'Hooft's $\eta$
symbols  \cite{t'hooft}~
\linebreak
$
\eta^{ab}_k=\epsilon_k{}^{ab0}+ {1 \over 2} \epsilon_{kij}
\epsilon^{i j a b}
$
 with    $i,j,k =\{ 1,2,3 \}$.
Some useful properties of $\eta_k^{ab}$  are  given  in Appendix I.
A point to notice is that
$M_4$ is an oriented K\"ahlarian manifold with $c_1(M_4)_R=0$
from  the postulate 1b. Thus at least  the
reduction of the structure group $SO(4) \rightarrow U(2)$ is possible
 for the  $\Lambda = 0$~case ( see Ref.~ 21 and Fig. 1).
\par
Furthermore we  assume  the parallelizability of $ \tilde  e^a_\nu$
 with  the Levi-Civita connection $\Gamma^\sigma_{\mu \nu}$ and
 the  frame connection $ \omega^{ab}_\mu$ defined by
$\tilde e^a_\mu$.
 This  is a
sufficient condition for the metricity of $\tilde g_{\mu \nu}~(~
\nabla_\tau \tilde g_{\mu \nu}=\partial_\tau \tilde g_{\mu \nu} -
\Gamma^\sigma_{\mu  \tau} \tilde  g_{\sigma \nu}-
\Gamma^\sigma_{\nu  \tau} \tilde g_{\sigma \mu}=0$~);
\begin{equation}
\nabla_\nu(\tilde e)\tilde  e_a^\mu = \partial_\nu \tilde e ^\mu_a
+\Gamma^\mu_{\sigma \nu}(\tilde e)\tilde e^\sigma_a
-\omega^b{}_{a \nu}(\tilde e)\tilde e^\mu_b = 0.
\label{eq:three}
\end{equation}
{}From this equation, the   relation between Riemannian
tensor and the curvature tensor  $F^k$ is given by
\begin{equation}
{}^{(-)} R_{\mu \nu}{}^{ \rho \tau}(\tilde g(\tilde  e))=
4 F_{\mu \nu}{}^{ab}(\tilde e)
                      \tilde   e_a^\rho \tilde e_b^\tau
= 4 F_{\mu \nu}{}^k(\tilde e)  \eta_k^{ab}
                      \tilde   e_a^\rho \tilde e_b^\tau, \
\label{eq:four}
\end {equation}
where
$
 R_{\mu \nu \rho \tau}= {}^{(+)} R_{\mu \nu \rho \tau}
+{}^{(-)} R_{\mu \nu \rho \tau}
$.
It is well known that
the  Riemann curvature tensors over $M_4$   are
written in block diagonal form of $6 \times 6 $ matrix \cite{besse};
\[
R= \pmatrix {
{}^{(+)} R_{\mu \nu \rho \tau}
 &
 \cr
{}^{(-)}R _{\mu \nu \rho \tau}
 &
 \cr
}
{}~
= \pmatrix {
{}^{(+)} W_{\mu \nu \rho \tau}
+ {}^{(+)} S_{\mu \nu \rho \tau}, &
 K_{\mu \nu \rho \tau} \cr
  {}^t K_{\mu \nu \rho \tau},
& {}^{(-)} W_{\mu \nu \rho \tau}
+ {}^{(-)} S_{\mu \nu \rho \tau}  \cr
}.
\]
${}^{(+)}W_{\mu \nu \rho \tau} $ is the self-dual part of the Weyl tensor and
${}^{(-)}W_{\mu \nu \rho \tau} $ is its anti-self-dual part.
${}^{(\pm)}S_{\mu \nu}{}^{ \rho \tau}  \propto
(\delta_{[  \mu}^{[\rho} \delta_{ \nu ]}^{\tau]}  \pm {1 \over 2} \epsilon
_{\mu \nu}{} ^{\rho \tau}) R
$~ and
$K_{\mu \nu}{}^{ \rho \tau}  \propto
\delta_{[  \mu}^{[\rho} \Phi_{ \nu ]}{}^{\tau]}$,
where $ \Phi_\mu{}^\tau= R_\mu{}^\tau -{1 \over 4} \delta_\mu^\tau R $
is the trace free part of the Ricci tensor and $R$ is the scalar curvature.
\\
Postulate~3~~ for $\Sigma^k$ field~:
\par
$ \{\Sigma^k \}$  are a trio of  $su(2)$ valued  2-forms.
 We suppose that  the index $k$ of  $\Sigma^k$ and $\omega^k$ denotes
the anti-self-dual part of  $so(4)$ index.  Namely $su(2)$ is a Lie algebra
of $ SU(2)_L$  which comes from $SPIN(4)= SU(2)_L \times SU(2)_R$
( the double covering group of  $SO(4)$).
\par
Our stance for this model is that the fundamental  variables are not
$\tilde e^a$  but ${\bf \omega^k} $ and
${\bf \Sigma^k}$. We seek the solutions  of them which satisfy the
above postulates, equations of  motion and the gauge fixing conditions.
These conditions specify the manifolds concerning to our model
 and the property of  metrics
or almost complex structures on them.
\par
We now turn to the gauge fixing conditions which we set to restrict
 five degrees of the freedom of $\Sigma^k$.
In this topological model, there exists a symmetry generated by a
parameter 1-form $\theta^k$ in addition to the $SU(2)_L$ (with a
$su(2)$ valued  0-form $\upsilon^k$) and diffeomorphism
(with a vector field $\xi^{\mu}$)
symmetries,
\begin{equation}
  \delta \omega^k_\tau
   =D_\tau\upsilon^k + ({\cal L}_\xi \omega^k)_\tau
    +{\Lambda \over 12}\theta_\tau^k  \ ,
\qquad
  \delta \Sigma^k_{\mu \nu}
  =2(\Sigma_{\mu \nu} \times \upsilon )^k +
({\cal L}_\xi \Sigma^{k})_{\mu\nu }
   +D_{[\mu}\theta^k_{\nu]}. \label{eq: six}
\end{equation}
 The $\theta^k$-symmetry is regarded as a `restricted' topological
symmetry which preserves the equations of motion (\ref{eqn:ins1})
 or (\ref{eqn:ins2}).
 With the appearance of the $\theta^k$-symmetry, the theory turns out
to be on-shell reducible in the sense that the transformation laws
(\ref{eq: six}) are invariant under
\begin{equation}
  \delta \upsilon^k = - \frac{\Lambda}{12} \epsilon^k
 +\rho^\sigma\omega^k_\sigma  \ ,
\qquad
  \delta \theta_{\mu}^k = D_{\mu} \epsilon^k
                      + 2  \rho^{\nu} \Sigma_{\nu \mu}^k \ ,
\qquad
    \delta \xi^{\mu} = - \rho^{\mu} \ ,\label{eq: seven}
\end{equation}
as long as the equations of  motion are satisfied.
 The transformations with parameters $\epsilon^k$ and $\rho^{\mu}$
correspond to redundant $SU(2)$ and redundant
diffeomorphism, respectively.
\par
 Our strategy to construct a topological quantum field theory is to
consider the following five  equations as  gauge fixing conditions for the
$\theta^k$-symmetry (except for the redundant part of the symmetry).
\\
Postulate~ 4~~ for the gauge fixing conditions~for the $\theta^k$-symmetry :
\begin{equation}
    {}^{t.f.} \Sigma^i \wedge \Sigma^j
    \equiv \Sigma^{(i} \wedge \Sigma^{j)}
    - {1 \over 3} \delta_{ij} \Sigma^k \wedge \Sigma_k = 0.
\label{eq:eight}
\end{equation}
 These   constraints were  imposed in the
original 2-form Einstein gravity  \cite{capovilla} and  are necessary
and sufficient conditions that $\Sigma^k$  comes from a   vierbein
$e^a= e^a_\mu dx^\mu$.
\begin{equation}
\Sigma^k(e) = - \eta^k_{ab}e^a \wedge e^b. \label{eq:nine}
\end{equation}
(We should remark that  $e^a$ is independent  of $\tilde e^a$
in this stage).
$\{ \Sigma^k(e) \}$  have 13 degrees of freedom.
The   Riemannian  metric  $g_{\mu \nu}= e^a_\mu  e^b_\nu \eta_{a b}$
can be  expressed  in terms of $\Sigma^k(e)$;
\begin{equation}
g^{1 \over 2} g_{\mu \nu}
           = -\frac{1}{12} \, {\epsilon^{\alpha \beta \gamma \delta}}
                             \, {\Sigma_{\mu \alpha}}_k
                             ({\Sigma_{\beta \gamma}} \times
                               {\Sigma_{\delta \nu}})^k \ ,
            \qquad g \equiv det(g_{\mu \nu})\ .
\label{eq:ten}
\end{equation}
 Such a 2-form $\Sigma^k(e) $ is anti-self-dual with respect to world
indices $\mu, \nu$ by the Hodge dual operation ~$\ast_{g(\Sigma(e))}$
 which is defined  via Eq. (\ref{eq:ten}).
( But it does not necessarily
mean that $\Sigma^k(e)$ belongs to   $su(2)$ valued
anti-self-dual 2-forms part  only because these eigenspaces of
the dual operation vary as the deformations of
$\Sigma^k$ and $g_{\mu \nu}(\Sigma)$.)
\par
 The set of equations  (\ref{eqn:ins1}) or (\ref{eqn:ins2})
and (\ref{eq:eight}) arose before as
an ansatz within the framework of 2-form Einstein gravity with the
cosmological constant  \cite{capovilla}${}^,$\cite{samuel}.
 We consider them to be {\it gravitational instanton equations}.
The  degrees of the freedom of the fundamental
fields are completely fixed by the above conditions (see Table 1).
\par
We also assume  the parallelizability of $e^a_\nu$
 with the Levi-Civita connection and   the frame connection defined by
$e^a_\nu$ as before. The equation
\linebreak
 $\nabla(e)_{[\mu}\nabla(e)_{\nu]} \Sigma^k_{\sigma \tau} =0$,
 which comes from   Eq. (\ref{eq:nine}) and this parallerizability
 yield   the following relation between
the curvature tensor $F^k(e)$ and Riemannian
curvature tensor;
\begin{equation}
{}^{(-)} R_{\mu \nu \rho \tau}(  e)= 4 F_{\mu \nu}{}^k(e)
                         \Sigma_{\rho \tau}{}_k(e).    \label{eq:twelve}
\end {equation}
\par
Now  we will explain
 what kinds of  Riemmanian tensors  are derived  from the solutions
 of  $\{\Sigma^k \}$
  which satisfy  Eqs. (\ref{eq:eight}),   (\ref{eq:twelve})
and (\ref{eqn:ins1})
 for $\Lambda \not= 0$
( or Eqs. (\ref{eq:eight}),   (\ref{eq:twelve})
 and (\ref{eqn:ins2}) for $\Lambda = 0$)
and show the definitions of each moduli space.
\\
\\
\par
(a) $\Lambda \not=0$ case~:~
\par
Using the property of   Riemannian manifold
such as  torsion tensor $T^a \propto  D(e)e^a =0$  and Eq.~(9)
 we obtain $D(e) \Sigma^k(e)=0$.
Comparing   this  with $D( \tilde e) \Sigma^k(e)=0$,
\begin{equation}
\omega^k (\tilde e)= \omega^k (e),~~F^k_{\mu\nu}(\tilde e)=
F^k_{\mu\nu}( e).
\label{eq:thirteen}
\end{equation}
Thus we obtain that
\begin{equation}
{}^{t.f.} F^i(e) \wedge F^j(e)
={}^{t.f.} F^i(\tilde e) \wedge F^j(\tilde e) = 0.
\label{eq:fourteen}
\end{equation}
which leads to~~ $\Sigma^k(e)=\Sigma^k(\tilde e)$.
(Note that    $\Sigma^k(e)=\Sigma^k(\tilde e)$ is not a
sufficient condition for
\par
By substituting
 $F^k(e)=F^k(\tilde e)={\Lambda \over 12} \Sigma^k( e)$
 into Eq. (\ref{eq:twelve})
\begin{equation}
{}^{(-)} R_{\mu \nu \rho \tau}(\Sigma(e))= {}^{(-)}
S_{\mu \nu \rho \tau}(\Sigma(e)).
\end{equation}
 Namely $M_4$ becomes a   conformally
self-dual Einstein manifold.
\begin{equation}
\Lambda \not=0 ;~~
R_{\mu \nu}(e) = \Lambda g_{\mu \nu}(e)~~ {\rm and}~~
{}^{(-)} W_{\mu \nu \rho \tau}(e) = 0.
\label{eq:fifteen}
\end{equation}
\par
The moduli space  in this case is defined by $\omega^k$ only ;
\begin{equation}
{\cal M}(\omega)=\{ \omega^k \mid \omega^k \in su(2) \otimes \wedge^1,~~
 {}^{t.f.} F^i \wedge F^j=0 \}
/\{SU(2) \times diffeo.\}.
\label{eq:sixteen}
\end{equation}
It corresponds to the moduli space of the  conformally self-dual Einstein
metrics because these metrics are represented by $\omega^k$
via Eqs.  (\ref{eqn:ins1}) and  (\ref{eq:ten}).
If we consider  only   compact Einstein conformally
self-dual Riemannian manifolds
with $R > 0 \ (\Lambda > 0)$, then $M_4$ is either isometric to $S^4$, or
to $CP^2$, with their standard metrics from  the theorem given by Hitchin
  \cite{besse}.
So  the solution $\Sigma^k(e)$  determines the standard metric on
$S^4$ or the Fubini-Study metric on $CP^2$ for $\Lambda > 0$.
\\
\par
(b) $\Lambda = 0$ case~:~
\par
 From Eqs. (\ref{eq:four}) and
(\ref{eqn:ins2}) ,
the Riemannian  tensor  defined by $\tilde e^a$ is self-dual~;~
\begin{equation}
{}^{(-)} R_{\mu \nu \rho \tau}(\tilde e)= 0
\end{equation}
so $(M_4, g)$ is a Ricci-flat K\"ahlerian manifolds.
The following theorem gives the characterization of Ricci-flat
K\"ahlerian manifolds:
\\
Theorem ( Hitchin \cite{besse})
\par
Let $M_4$ be a compact connected oriented Riemannian manifold.
If $M_4$ is Ricci-flat and ${}^{(+)}W_{\mu \nu \tau \rho}=0$,
then  $(M_4, g)$ is  one of the following four cases~:~
\\
(1)~ $(M_4, g)$ is  flat, i.e.   is covered by a flat  4-torus
\\
(2)~ $(M_4, g)$ is a K\"ahler-Einstein K3-surface ($\pi_1=1$)
\\
(3)~ $(M_4, g)$ is a K\"ahler-Einstein  Enriques surface ($\pi_1= Z_2$ )
\\
(4)~ $(M_4, g)$ is  the  quotient of a
K\"ahler-Einstein Enriques surface by a free anti-holomorphic isometric
involution ($\pi_1=Z_2 \times Z_2$).
\par
(We   should better take
the opposite orientation of  $M_4$  and replace
$\eta^k_{ab}$ by  self-dual notation  $\bar \eta^k_{ab}$
 for the  $\Lambda=0$ case
 so that  ${}^{(+)}R_{\mu \nu \rho \tau}(e)=0$ and
Einstein- K\"ahler forms $\{\Sigma\}$
belong to  self-dual  $(1,1)$ form. )
\par
 The relation between
the vanishing covariant derivative with a Levi-Civita connection
 and the holonomy group $Hol(g)$
asserts that
$Hol(g) \subseteq U(2)$ for K\"ahlerian manifolds
with the complex dimension two.
These compact K\"ahlerian manifolds  with $c _1(M_4)_R=0$
are exactly the compact complex manifolds admitting
a K\"ahler metric with zero Ricci form (or
equivalently the compact complex manifolds
with restricted holonomy group contained in the
special unitary group ).
We now investigate the properties  of the  metrics or the
complex structures defined by
$\Sigma^k(e)$ on these manifolds. We divide  these manifolds into two
groups.
\\
\\
case~b-1~( when the canonical line bundle $K$ is trivial)~:~
\\
$M_4$ is a $K3$-surface  or a four-torus  $T^4$.
\par
On these two manifolds, the following reductions of $P_{U(2)}$
 are  possible due to the fact that  the  canonical line bundles $K$  over them
are trivial.
Actually,   the restricted holonomy group  $Hol_0(g)$ reduces to
the identity exactly when a  metric is flat for  $T^4$.  Though  $T^4$
is not a simply-connected,  $Hol(g)= Hol_0(g)$ happens. Therefore
the reduction  $P_{U(2)} \rightarrow P_I$  is possible    and all
  frame connections can  be gauged away when metric is flat
\cite{besse}.
A $K3-$ surface is, by definition,
a compact  simply-connected complex surface
with trivial canonical line bundle $K$ and $b_1=0$.
For a Calabi-Yau metric (a K\"ahler-Einstein metric)
$Hol(g)= Hol_0(g) \subseteq SU(2)_{\rm R} \cong SP(1)$.
Thus   $SU(2)_{\rm L}$  connections
$\{\omega^k(\tilde e)\} $ can be gauged away.
\par
 In these cases   $D(\tilde e) \Sigma^k(e)=0$
reduces to $d\Sigma^k(e)=0$ for all $k$.
{}From  $d\Sigma^k(e)=0$ and $D (e) \Sigma^k(e)=0$,
we obtain  $\omega^k(e)= 0$ and  the Ricci-flat K\"ahler metric.
\begin{equation}
\Lambda= 0~ ;~~R_{\mu \nu }(e)=0,~~
{}^{(+)}W_{\mu \nu \tau \sigma}(e) = 0,~~ \omega^k(e) = 0 .
\label{eq:seventeen}
\end{equation}
$\{ \Sigma^k(e) \}$ define a  Calabi-Yau metric on a $K3$-surface
  or a flat metric on $T^4$.
The gravitational instanton equations for a $K3$-surface
with Calabi-Yau metrics and a  four-torus with flat metrics reduce to
\begin{equation}
      d\Sigma^k=0 \ ,
\qquad \qquad
      {}^{t.f.} \Sigma^i \wedge
\Sigma^j =0 \ .
                                               \label{eq:eighteen}
\end{equation}
 These equations give the Ricci-flat condition and  restrict
$\{\Sigma^k \}$
to be a trio of closed Einstein-K\"ahler forms
 \cite{kunitomo}${}^,$\cite{capovilla}.
 In Ref. 24, Plebanski used these equations to
derive his `heavenly equations'.
 A manifold which satisfies Eq.~(\ref{eq:eighteen}) is called
hyperk\"ahlerian.
On hyperk\"ahlerian manifolds, a trio of K\"ahler forms
$\{ \Sigma^k(e)\}$ is represented by
\begin{equation}
 \Sigma^k(e)  = - \eta^k_{ab} e^a \wedge e^b \propto
g_{\alpha \bar \beta}  J^{k \bar \beta }_{\bar \gamma} dz
^\alpha \wedge d\bar z^ {\bar  \gamma},
\end{equation}
where
$\{ J^k \}$  are  a trio of the $g$-orthogonal  complex structures
which  satisfy the  quaternionic relations and  $g_{\alpha \bar \beta}$ is an
Hermite  symmetric metric. $z$ and $\bar z$  denote  complex local
coordinates on these manifolds.
\par
The  moduli space is
the equivalent class of  a trio of the  Einstein-K\"ahler forms
(the hyperk\"ahler forms) $\{\Sigma^k(e) \}$~:~
\begin{equation}
{\cal M}(\Sigma)=\{ \Sigma^k \mid \Sigma^k \in su(2) \otimes \wedge^2~~,
 {}^{t.f.} \Sigma^i \wedge \Sigma^j=0,~d\Sigma^k=0  \}
/ \{diffeo.\}. \label{eq:nineteen}
\end{equation}
\\
\\
case~b-2~ (when  the canonical line bundle $K$ is not trivial)~:~
\\
$M_4$ is $K3/Z_2$,  $K3/{Z_2 \times Z_2}$, or $T^4/\Gamma$ where
$\Gamma$ is  some discrete group.
\par
In these cases, the reductions of $P_{U(2)} \rightarrow P_{{SU(2)}_R}$
  are not possible because the canonical line bundles  are not trivial.
{}From Eqs. (\ref{eqn:ins2}), (\ref{eq:eight}) and (\ref{eq:twelve}),
the Riemannian  self-dual tensors are also derived;
\begin{equation}
\Lambda=0 ;~~R_{\nu \mu}(e)=0~~,
{}^{(+)}W_{\mu \nu \tau \sigma}(e) = 0.
\end{equation}
The Hitchin's theorem states that $\{\Sigma^k(e)\}$ form
Einstein-K\"ahler metrics on these manifolds.
In these cases
$Hol_0(g) \subset SU(2)_{\rm R}$  but
$Hol_0(g) \not= Hol(g) $ is held.
They are called as the locally hyperk\"ahlerian K\"ahlerian manifolds
and some informations from  $\omega^k$ and $\Sigma^k$    will
be needed to describe the moduli spaces.
\newpage
\bf{III.  BRST SYMMETRY  }
\par
In this  section we will explain the BRST symmetry of the model
in the $\Lambda \not= 0$  case.
Our action is invariant under the usual gravitational  transformations,
the restricted topological  transformations.
These transformations are invariant under the   redundant
 transformations of them.
 We shall denote the  BRST versions
of the gravitational transf. $\delta^G_{\tilde c} $,
  the restricted topological transf. $\delta^S$  and
the redundant transf. $\delta^G_{\tilde\gamma}$, respectively.
\par
( For $\Lambda =0$ this model belongs to  the BF- type model
so more careful investigations  into the symmetries is necessary
 \cite{mabe}. )
We introduce the following notations for the  BRST symmetry~;~
\par
\begin{center}
\begin{tabular}{lcr}
\multicolumn{3}{l} {\bf diffeo.~$ \times$~ SU(2)~$\rightarrow$~BRST  }
\\
\multicolumn{1}{r}{ghost}
&anti-ghost
&N-L~ field
\\
\it
$(\xi^\nu, \upsilon^k) \rightarrow \tilde c^k \equiv (c^\nu, c^k)$
&$\tilde b^k \equiv (b_\nu dx^\nu, b^k)$
&$\tilde \pi^k \equiv (\pi_\nu dx^\nu, \pi^k)$
\end{tabular}
\end{center}
\begin{center}
\begin{tabular}{lcr}
\multicolumn{3}{l} {\bf redundant~diffeo.~$ \times$
redundant~ SU(2)~$\rightarrow$~BRST}
\\
\multicolumn{1}{r}{ghost}
&anti-ghost
&N-L~ field
\\
\it ~~~~~
$(\rho^\nu, \epsilon^k) \rightarrow
\tilde \gamma^k \equiv (\gamma^\nu,  \gamma^k)$
&$\tilde \beta^k \equiv (\beta_\nu dx^\nu, \beta^k)$
&$\tilde \tau^k \equiv (\tau_\nu dx^\nu, \tau^k)$
\\
\\
\multicolumn{3}{l} {\bf restricted~ topological~sym.  ~$\rightarrow$~BRST}
\\
\multicolumn{1}{r}{ghost}
&anti-ghost
&N-L~ field
\\
\it ~~~~~~~~~~~~~~~~~~~~
$\theta^i \rightarrow
 \phi^i$
&$\chi^{ij}$
&$\pi^{ij}$
\end{tabular}
\end{center}
\par
The on-shell BRST  transformations of this model  are given by
\begin{eqnarray}
(1)~ \delta_B \omega^i_\mu~ &=& D_\mu c^i+ ({\cal  L}_c \omega^i)_\mu+
{ \Lambda \over 12} \phi^i_\mu \nonumber \\
&\equiv& \delta^G_{ \tilde c}
\omega^i_\mu +\delta^S\omega^i_\mu, \nonumber
\\
(2)~ \delta_B \Sigma^i_{\mu\nu} &=& 2(\Sigma_{\mu\nu} \times  c)^i+
({\cal  L}_c \Sigma^i)_{[\mu \nu]}+ D_{[\mu} \phi^i_{\nu]}
\nonumber \\
&\equiv& \delta^G_{\tilde c} \Sigma^i_{\mu\nu}+ \delta^S\Sigma^i_{\mu\nu},
\nonumber \\
(3)~~ \delta_B c^i~~&=& - (c \times  c)^i+ {\cal  L}_c c^i
-{ \Lambda \over 12} \gamma^i + \gamma^\sigma \omega^i_\sigma
\nonumber \\
&\equiv& - (c \times  c)^i+ {\cal  L}_c c^i
 + \hat \gamma^i, \nonumber \\
(4)~~ \delta_B b^i~~ &=& - 2 (b \times  c)^i
+ {\cal  L}_c b^i
+ \pi^i,
\nonumber \\
(5)~~ \delta_B \pi^i~~ &=&  2 (\pi \times  c)^i
+ {\cal  L}_c \pi^i
+{ \Lambda \over 6}( b \times \gamma)^i
+  \gamma^\sigma D_{\sigma}b^i,
\nonumber \\
(6)~~ \delta_B c^\mu~~ &=&  c^\rho \partial _\rho c^\mu- \gamma^\mu,  \nonumber
\\
(7)~~ \delta_B b_\mu~~ &=&  ({\cal  L}_c b)_\mu +  \pi_\mu,
\nonumber \\
(8)~~ \delta_B \pi_\mu~~ &=&  ({\cal  L}_c \pi)_\mu +
({\cal  L}_\gamma b)_\mu,
\nonumber \\
(9)~~ \delta_B \gamma^i~~ &=& 2(\gamma \times  c)^i + {\cal  L}_c \gamma^i+
\gamma^\sigma \phi^i_\sigma
\nonumber \\
&\Rightarrow&  \delta_B \hat  \gamma^i = 2( \hat \gamma \times  c)^i +
 {\cal  L}_c \hat \gamma^i,
 \nonumber \\
(10)~~ \delta_B \beta^i~~ &=& 2 (\beta \times  c)^i
+ {\cal  L}_c \beta^i
+ \tau^i,
\nonumber \\
(11)~~ \delta_B \tau^i~~ &=&- 2 (\tau \times  c)^i
+ {\cal  L}_c \tau^i
+{ \Lambda \over 6}( \beta^i \times \gamma)^i
+  \gamma^\sigma D_{\sigma}\beta^i,
\nonumber \\
(12)~~ \delta_B \gamma^\mu~~ &=&  {\cal  L}_c \gamma ^\mu, \nonumber \\
(13)~~ \delta_B \beta_\mu~~ &=&  ({\cal  L}_c \beta)_\mu+ \tau_\mu,
\nonumber \\
(14)~~ \delta_B \tau_\mu~~ &=&  ({\cal  L}_c \tau)_\mu +
({\cal  L}_\gamma \beta)_\mu,
\nonumber \\
(15)~~ \delta_B \phi^i_\mu~ & =& - 2(\phi_\mu \times  c)^i
+ ({\cal  L}_c \phi^i)_\mu
+ D_\mu \gamma^i
+ 2 \gamma^\sigma \Sigma^i_{\sigma \mu} \nonumber
\\
&\equiv& \delta^G_{\tilde c} \phi^i_\mu + \delta^G_{\tilde \gamma}
\omega^i_\mu +{24 \over \Lambda}
\gamma^\rho (F^i_{\mu \rho}-{\Lambda \over 12 }
\Sigma^i_{\mu \rho}), \nonumber \\
(16)~~\delta_B \chi^{ij}~~ &=&  2 (\chi\times  c)^{ij}
+ {\cal  L}_c \chi^{ij}
+{ \Lambda \over 6}( \chi\times \gamma)^{ij}
+  \gamma^\sigma D_{\sigma}\chi^{ ij},
\nonumber \\
(17)~~ \delta_B \pi^{ij}~~ &=& \chi^{ij} - 2 (\pi \times  c)^{ij}
+ ({\cal  L}_c \pi^{ij}),
\label{eq:twoone}
\end{eqnarray}
where $\delta^G_{\tilde \gamma} \omega^i_\mu$ denote
the redundant  transformations and are given by the
replacement of the parameters  $\tilde c \rightarrow \tilde \gamma$.
The characteristic feature  of  our BRST symmetries is  the
 presence of the restricted topological
symmetry of the fundamental fields~;
  $\delta^S \omega^i_\mu= { \Lambda \over 12}
\phi^i_\mu,~\delta^S \Sigma^i_{\mu\nu}=
D_{[\mu} \phi^i_{\nu]}$.
\par
  As already pointed out,  this action  comes  to a
Witten-type action  for  $\Lambda \not= 0$ under
$\alpha \rightarrow 0$ limit by removing $\Sigma^k$ using the
equation of motion.
The restricted topological symmetry
 can be interpreted as the supersymmetry  for a Witten- type model.
The supersymmetric pair
$(\delta \omega^k_\mu, \phi^k_\mu)$ is  important
because  it forms a  basis of the tangent space of the
moduli space while  the other  pair  $(\delta\Sigma^k_{\mu\nu},
D_{[\mu}\phi^k_{\nu]})$ is
the auxiliary one and can be removed by using the equation of
motion.
The symmetries in the $\Lambda \not= 0$ case  are interpretable as
\begin{equation}
 \{ SU(2) \times {\rm diffeo.} \times {\rm super~sym. } \}
/\{ {\rm redundant}~ SU(2) \times {\rm redundant~ diffeo.}\}.
\end{equation}
 The   transformations of these fields also  end in
those   of the ordinary  Witten-type theory    even for  off-shell
except $\phi^k_\mu$ and
$\Sigma^k_{\mu \nu}$ by redefining the redundant $SU(2)$ ghost
 as $\hat \gamma^i \equiv-{\Lambda \over 12}
\gamma^i+\gamma^\mu \omega_\mu^i$.
The BRST symmetry of $\phi^i$ agrees with Witten-type  theory on-shell.
Thus this model coincides with the Witten -type topological
gravity  model given by Torre \cite{torre}
up to the secondary Chern number
which is our classical action after eliminating $\Sigma^k$
under $\alpha \rightarrow 0$ limit.
\par
{}From now on, we  will replace Lie-derivative
${\cal  L}_c $ ($ {\cal L}_\gamma$ )
with the modified one
$
\tilde {\cal L}_c \omega^k_\mu \equiv  ({\cal L}_c \omega^k)_\mu
 - D(c^\mu \omega^k_\mu)
$~
($
\tilde {\cal L}_\gamma \omega^k_\mu
\equiv ({\cal L}_\gamma \omega^k)_\mu
 - D(\gamma^\mu \omega^k_\mu)
$)~
so that $\delta_B \omega^k_\mu$ ($\delta_B \phi^k_\mu$ )
 still remains  in $P_{SU(2)} \times Ad su(2) \otimes \wedge^1$,
\begin{equation}
(\tilde {\cal L}_c\omega^i)_\mu=2c^\tau F^i_{[\tau \mu]},~
{}~
\tilde {\cal L}_c\Sigma^i_{\mu \nu}
=2 D_{[\mu}c^\sigma \Sigma^i_{\sigma\nu]}+
3c^\sigma D_{[\mu}
\Sigma^i_{\sigma\nu]}.
\end{equation}
\par
  Before  we  proceed, it will be useful to introduce
general spin bundles $\Omega^{m,n}$, the space of fields with spin
$(m,n)$ of SU(2)$_L$ $\times$ SU(2)$_R$  \cite{penrose}.
Let us denote by $\Omega^{1,0} $, $\Omega^{0, 1} $ the two complex vector
bundles on $M_4$ associated  with the defining 2-dimensional representations of
the
two factors.  These will only exist globally if $M_4$ is a spin manifold , i.e.
the 2nd Stiefel - Whitney class $w_2(M_4)=0$.
Let us denote  by
$\Omega^{m,n} \equiv  S^m \Omega^{1,0} \otimes S^n \Omega^{0,1}$
the tensor product of the $m$-th symmetric power bundle of
$\Omega^{1,0}$ and
the $n$-th symmetric power bundle of
$\Omega^{0,1}$.
For example, the space of $P_{SU(2)} \times Ad su(2) $ valued 1-forms
$\delta \omega^k$ is
$\Omega^{2,0} \otimes \Lambda^1 \simeq \Omega^{2,0}
\otimes \Omega^{1,1}$ while the space  of
$(\xi^{\mu}, \upsilon^k)$
is equivalent to $ \Lambda^1 \oplus \Omega^{2,0}  \simeq
 \Omega^{1,1}  \oplus \Omega^{2,0} $ (see Table 2).
\newpage
\bf{ IV.  ZERO MODES  IN THE  $\Lambda \not= 0$ CASE }
\par  To  know the number of zero modes in the quantum action $S_q$, we
consider the moduli space ${\cal M}$ defined by our instanton equations
Eq.~(\ref{eq:sixteen})   for  conformally
self-dual Einstein manifolds.
\begin{equation}
  {\cal M}(\omega)
  = \{  \omega_0^k \vert  \omega^k \in  su(2) \otimes
\Omega^{1,1},~~
    {}^{t.f.} F^i \wedge F^j =0  \} /
             \{ SU(2) \times diffeo. \}  \ .
                                                  \label{eq:fione}
\end{equation}
 Given a solution $(\Sigma^k_0, \omega^k_0)$ of the instanton equations,
the tangent space $T({\cal M})$ of ${\cal M}$ is
 the space of infinitesimal deformations
$\delta  \omega^k$
 which satisfy linearized instanton equations
modulo  deformations generated by SU(2) ( the subgroup of SO(4))
transformation and diffeomorphism:
\begin{equation}
  T({\cal M}(\omega))
  = \{  \delta\omega^k \vert \delta\omega^k \in \Omega^{2,0} \otimes
\Omega^{1,1}, D_1  \delta\omega^k= 0 \} /
            \{ SU(2) \times diffeo. \}  \ .
                                                  \label{eq:fitwo}
\end{equation}
where
\begin{equation}
 D_1  \delta \omega^k \equiv
           \, {}^{t.f.} F^i_0 \wedge  D \delta \omega^j  = 0 .
                                           \label{eq:fithree}
\end{equation}
This linearized instanton equation  is  derived by substituting equation
$ D \delta\omega^k- {\Lambda \over 12} \delta \Sigma^k=0$  into
${}^{t.f.} \Sigma_0^i \wedge \delta \Sigma^j = 0$.
\par
 We define the following sequence of mappings on a compact conformally
self-dual Einstein manifold in terms of the spin bundles:
\begin{equation}
   0 \stackrel{D_{-1}} \to \
C^\infty( {  \Omega^{1, 1} \oplus  \Omega^{2, 0}} )\
   \stackrel{D_0} \to \
C^\infty ({\Omega^{2, 0} \otimes \Omega^{1, 1}}) \
   \stackrel{D_1} \to \
C^\infty({\Omega^{4, 0}}) \
   \stackrel{D_2} \to \ 0 \ ,
                                             \label{eq:fifour}
\end{equation}
where the symbol sequence is
\begin{eqnarray}
  &~~~~~  V_0 ~~~~~~~~~~~~~   V_1  ~~~~~~~~~ V_2 &  \nonumber
\\
 0 \rightarrow
&\Omega^{1, 1} \oplus  \Omega^{2, 0}
   \rightarrow
 \Omega^{2, 0} \otimes \Omega^{1, 1}
   \rightarrow
 \Omega^{4, 0} &
   \rightarrow 0  \nonumber .
                                             \end{eqnarray}
 In the above sequence $D_{-1}$ and $D_{2}$ are identically zero
operators.
 The operator $D_0$ is defined by
\begin{equation}
  D_0 (\xi^{\mu},\upsilon^k ) \equiv
{\tilde {\cal L}}_\xi \omega^k   +  D\upsilon^k    \ .
                                             \label{eq:fifive}
\end{equation}
 We can easily check the ellipticity of the deformation complex.
 Defining the inner product in each space $V_i$, we can introduce
the adjoint operators $D_0^*$ and $D_1^*$ for $D_0^{}$
and $D_1^{}$ respectively and the Laplacians $\triangle_i$;
$ \triangle_0 = D_0^* D_0^{} \, , \
  \triangle_1 = D_0^{} D_0^* + D_1^* D_1^{} \, , \
  \triangle_2 = D_1^{} D_1^* \,$.
 We may then define the cohomology group on each $V_i$,
\begin{equation}
 H^i \equiv {\rm Ker}\, D_i/{\rm Im}\, D_{i-1} \ .
                                                \label{eq:fisix}
\end{equation}
 It is easy to show that $H^i$ is equivalent to the kernel of
$\triangle_i$, the harmonic subspace of $V_i$.
 These cohomology groups are finite-dimensional. We call
 the  dimension of  $ H^i$  $h^i$.
 The $H^1$ is exactly identical with the tangent space of
${\cal M}(\omega)$ in
Eq.~(\ref{eq:fitwo}), the dimension of which we need to know.
 On the space $V_0$, $H^0$ is equal to ${\rm Ker}\, D_0$ because
the image of $D_{-1}$ is trivial.
 In the $\Lambda \not = 0$ case, Torre found that ${\rm Ker}\, D_0$ is
equivalent to the space of the Killing vectors  \cite{torre}.
 The kernel of $D_2$ is the whole of the space $V_2$.
 Hence $H^2$ is the subspace of $V_2$ orthogonal to the mapping
$D_1$, or equivalently it is the kernel of $D_1^*$.
 The index of the elliptic complex is defined as the alternating sum,
\begin{equation}
  {\rm Index} \equiv h^0 - h^1 + h^2 \ .
                                              \label{eq:fiseven}
\end{equation}
  By applying the Atiyah-Singer index theorem \cite{shanahan}
to the elliptic complex,
we obtain
\begin{eqnarray}
 {\rm Index}&=&\int_{M_4}
 \frac{{\rm ch} ( \Omega^{2,0} \oplus  \Omega^{1,1} \, \ominus \,
                  \Omega^{2,0} \otimes \Omega^{1,1} \, \oplus \,
                  \Omega^{4,0} )\,
      {\rm td} (TM_4 \otimes {\bf C} ) }
      {{\rm e} (TM_4) } \\ \nonumber
&=& \int_{M_4}
 \frac{{\rm ch} ( \Omega^{2,0}
                  \ominus \Omega^{3,1}
                  \oplus \Omega^{4,0} )\,
      {\rm td} (TM_4 \otimes {\bf C} ) }
      {{\rm e} (TM_4) } \\ \nonumber
&=& 5 \chi - 7 \tau ,                \label{eq:fff}
\end{eqnarray}
where ${\rm ch},$ ${\rm e}$ and ${\rm td}$ are the Chern
character, Euler class and Todd class of the various vector bundles
involved.
 Therefore the alternating sum of $h^i$ in Eq.~(\ref{eq:fiseven}) is
determined by the Euler number $\chi$ and Hirzebruch signature $\tau$.
 By changing $\tau \rightarrow \mid \tau \mid $, this index can also be
 adopted to manifolds with the opposite orientation.
 \par
If $\Lambda > 0$, $h^1$ and $h^2$ are found to be zero  as shown
 by Torre  \cite{torre}.
The result  such as $h^1=0$ for the $\Lambda > 0$ case   agrees with
the one of Perry and Teo.  They showed that
the dimension of the    moduli space of
conformally self-dual metrics is zero on $S^4$ or $CP^2$ by
using the deformation complex for  the metrics \cite{perry}.
 Therefore from Eqs.~(\ref{eq:fiseven}) and  (33),    the
dimension $h^0$ is equal to the index,
\begin{equation}
  h^0 = 5 \chi - 7 \tau \ ,
\qquad
  h^1 = h^2 = 0 \ .
                                             \label{eq:finine}
\end{equation}
 The value of $h^0$, the dimension of the Killing vector space, agrees
with that obtained by a different method in Ref. 23.
For $S^4$ with the  standard metric, the dimension of the  isometry
 is given by dim.  $SO(5)= 10$, which  coincides with
$h^0=5 \chi-7\tau \mid_{\tau=0, \chi=2}=10$.
For $CP^2$ with the Fubini-Study metric, the dimension of the  isometry  is
given by dim.  $SU(3)=8$  which agrees  with
$h^0=5 \chi-7\tau \mid_{\tau=1, \chi=3}=8$.
 If $\Lambda < 0$, $h^0$ becomes zero  \cite{torre},
although $h^1$ and $h^2$ are not completely determined;
\begin{equation}
 h^0 = 0 \ ,
\qquad
 h^2-h^1 = 5 \chi - 7 \tau \ .
                                              \label{eq:fiten}
\end{equation}
\par
For   conformally self-dual Einstein  manifolds with $\Lambda < 0 $,
there are  two known examples, which  are
$  {\rm hyperbolic~surface} / \Gamma $
and ${\rm  boundary~ domain} / \Gamma$ where $\Gamma$ is  some  discrete
subgroup.
The dimensions of
their  moduli spaces of  conformally self-dual Einstein metrics are zero
due to the Mostov's rigidity \cite{mostov}.
Thus in these cases the dimensions of the moduli spaces of the
anti-self-dual frame  connections are also zero.
\newpage
\bf{V. THE PARTITION FUNCTION IN THE $\Lambda \not= 0$ CASE}
\par
For the purpose of the calculation of the partition function,
we first decompose  $\omega^k$
and $\Sigma^k$ fields as follows;
\begin{eqnarray}
&\omega^k = \omega_0^k + \delta\omega^k,
\\ \nonumber
&\Sigma^k = \Sigma_0^k + \delta\Sigma^k,   \label{eq:fone}
\end{eqnarray}
where $\omega_0$  and $\Sigma_0^k$ are the
background solutions of   conformally self-dual Einstein manifolds.
 $\delta\omega^k$ and $\delta\Sigma^k$ are
quantum fluctuations (=infinitesimal deformations).
\par
The BRST quantization of the Witten-type topological gravity model
in the  $\Lambda \not=$ 0 case is straightforward.
Twelve gauge fixing conditions for the  super symmetry
and seven ones for  SU(2) $\times$   diffeo.  are imposed.
The  gauge fixing conditions  for the  super symmetry
  consist of five gauge  fixing conditions for
the super  symmetry   except for the redundancy
and seven fixing  conditions  to remove the freedom
of the redundant symmetries.
We are fixing the gauge to be $  D_0 ^\ast \delta \omega^k =0$ for the
diffeomorphism and $SU(2)$
and   $  D_0 ^\ast \phi^k =0$ for the redundant
diffeomorphism and redundant $SU(2)$ ;
\begin{eqnarray}
D_0^\ast \delta \omega^k \equiv & (\tilde{\cal L}_c^\ast
\delta \omega^k, D^ \ast
\delta\omega^k)=0,
\nonumber  \\
&{\rm diffeo.} ~~~~~~{\rm SU(2)} \label{eq:ftwo}
\\
D_0^\ast \phi^k \equiv & (\tilde{\cal L}_\gamma^\ast  \phi^k,
D^ \ast \phi^k)=0,  \nonumber \\
&{\rm  red.~diffeo.} ~~~~~~{\rm red.~ SU(2)} \label{eq:fthree}
\\
(D_1 \delta \omega)^{ij} \equiv & {}^{t.f.} F^i \wedge D \delta \omega^j =0,
\nonumber \\
&{\rm  super } /\{{\rm red.~ diffeo. } \times {\rm red.}~ SU(2) \}
\label{eq:ffthree}
\end{eqnarray}
where $\ast$  denotes the  Hodge star dual operation and
${\cal O}^\ast \equiv - \ast {\cal O} \ast $  is  the  adjoint operator
of ${\cal O}$.
The operator  $D_0^\ast ~:~ C^\infty (\Omega^{2,0} \otimes \Lambda^1)
\rightarrow C^\infty (\Omega^{2,0} \otimes \Lambda^0)$ is the
adjoint operator of $D_0$~:~$\tilde c^k \rightarrow \delta \omega^k$,
\begin{equation}
D_0 \tilde c^k(\omega) \equiv
\tilde {\cal L}_c  \omega^k+ D_\tau c^k dx^\tau=
 2c^\nu D_{[\nu}
\omega^k_{\tau]} dx^\tau + D_\tau c^k dx^\tau,
\label{eq:ffour}
\end{equation}
where  $D=d+(\omega \times \cdot )$.
 The elements of the image of
$D_1~:~ C^\infty (\Omega^{2,0} \otimes \Lambda^1)
\rightarrow C^\infty (\Omega^{4,0} \otimes \Lambda^4 )$~
are 4-forms with symmetric
trace-free SU(2) indices, i.e.  sections of ~
 $\Omega^{4,0} \otimes \Lambda^4 \simeq \Omega^{4,0}
\otimes \Lambda^0 $.
\par
The gauge fixed quantum action  is  given by
\begin{equation}
S_q=S_{TG}+ \int \delta_B[\chi_{ij} (D_1 \delta\omega)^{ij} +
\tilde b_k \ast D_0^ \ast \delta\omega^k
+ \tilde \beta_k \ast D_0^\ast \phi^k  ]. \label{eq:ffive}
\end{equation}
When expanded out by using the properties of $\delta_B$ in
Eq. (\ref{eq:twoone}),
Eq. (\ref{eq:ffive})  reads
\begin{eqnarray}
S_q &=& S_{TG} +  \nonumber
\\  &+ \int& \pi_{ij}  (  D_1 \delta\omega )^{ij}
     + \tilde   \pi_k \ast   D_0^\ast \delta\omega^k
+  \tilde b_k \ast D_0^\ast D_0 \tilde c^k (\omega)
     +   \tilde b_k \ast D_0^ \ast  \phi^k  \nonumber
\\  &+ \int&  \chi_{ij}   (D_1 \phi)^{ij}
     +   \tilde \tau_k \ast  D_0^\ast \phi^k
+ \tilde  \beta_k  \ast  D_0^\ast D_0 \tilde \gamma^k(\omega)
     +    \tilde  \beta_k \ast D_0^\ast D_0 \tilde c^k( \phi) \nonumber
\\  &+ {} &  {\rm  other~ higher ~order~terms}.
\label{eq:fsix}
\end{eqnarray}
We are now ready to evaluate the partition function.
\begin{equation}
Z= \int  {\cal D}X (-S_q),
\label{eq:fseven}
\end{equation}
where ${\cal D}X$ represents the path integral over the
 fields such as $\delta\Sigma^k, \delta\omega^k$,
ghosts, anti-ghosts, N-L fields, etc.
The Gaussian integrals over the commuting
$ \tilde \beta - \tilde \gamma$
set of fields in Eq.~(\ref{eq:fsix})
 yields the determinant
$({\rm det} \Delta_0)^{-1}$ which cancels
with the ${\rm det} \Delta_0$ contribution  coming from
the anti-commuting set of fields  $ \tilde b- \tilde c$ set.
The term~
$
\tilde  of  \beta_k \ast D_0^\ast D_0 \tilde c^k( \phi)
$  is three-point interaction of  ghosts so
 does not contribute to the partition function.
\par
Consider now two terms $\tilde \tau_k \ast  D_0^\ast
\phi^k+\chi_{ij}(D_1\phi)^{ij}$~
(we absorb  the $ \tilde b_k \ast  D_0^\ast \phi^k$ term into
the $\tilde \tau_k \ast D_0^\ast \phi^k$ term).
In calculating their determinants we use a differential operator
$T=D_0^\ast \oplus  D_1$ and its adjoint operator $T^\ast$ ;
\begin{eqnarray}
{}~~&T& \nonumber \\ T^\ast T~
;~\Omega^{3,1} \oplus\Omega^{1,1} &\rightleftharpoons &
\Omega^{0,0} \oplus\Omega^{2,0}\oplus\Omega^{4,0} \nonumber \\
&T^\ast& \label{eq:feight}
\end{eqnarray}
One could show ${\rm det}T={\rm det}^{1 \over 2} (T^\ast T)
=({\rm det}\Delta_1)^{1 \over 2}$ by using matrix notations for
$T$ and $T^\ast$.
The $ \tilde \pi_k-\delta\omega^k-\pi_{ij}$ system of
commuting fields gives
 $ ({\rm det} \Delta_1) ^{- {1\over 2}} $ which cancels
with the determinant
$ \tilde \tau_k-\phi^k-\chi_{ij}$ system of anti-commuting fields.
\par
Since  the moduli space  has a vanishing dimension $h^1 =0 $
for the $\Lambda > 0 $ case, it consists of isolated points
such as   $CP^2$ with the Fubini-Study metric or $S^4$
with the standard metric.
We can write the partition function as
\begin{equation}
Z= \Sigma_{\rm instanton} ({\rm det} \Delta_1)^{-{1 \over 2}}
({\rm det} \Delta_0)^{-1} ({\rm det} \Delta_1)^{1 \over 2}
({\rm  det} \Delta_0)=\Sigma_{\rm instanton} \pm 1 = \pm 1
\label{eq:fnine}
\end{equation}
 up to  the secondary Chern class by projecting out
$h^0$ zero modes.
\par
We  comment the BRST symmetry of the $\Lambda=0$ case briefly.
Substituting $\Lambda=0$ in Eq.~(\ref{eq:twoone}) the
restricted topological symmetry
of the fundamental fields are   $\delta^S \omega^k_\mu=0, ~
\delta^S \Sigma^k_{\mu\nu}= D_{[\mu}\phi^k_{\nu]}$.
 The  difference is apparent  from the Witten type topological model
given by Kunitomo \cite{kunitomo},
 which has the following transformation for the fundamental fields~
$\delta^S \omega^k_\mu= \hat \phi^k_\mu, ~
\delta^S \Sigma^k_{\mu\nu}= \Psi^k_{[\mu\nu]}$.
The fermionic ghost zero modes for  $\phi^k_\mu$ of our model
are contained in the basis of the tangent space of
the  moduli space and further investigation is necessary to know
the precise value of the dimension.
\par
However on  a $K3$-surface and $T^4$, the special
situation occurs ~;~
$\omega^k$ can be gauged away  for on-shell
 and the  topological symmetry    reduces to
$\delta^S \Sigma^k= d\phi^k$.
The number of the free parameters of this
symmetry is not 12 but 9 due to the redundancy of
 $\{d^2 \phi=0 / d^3 \phi=0\} $. There is no need to
fix  $SU(2)$ and  redundant $SU(2)$ in this case.
We only fix  five degrees of the
 restricted topological symmetry and  four degrees of
diffeomorphism and four degrees of the
redundant diffeomorphism \cite{mabe}.
\newpage
\bf {VI. THE  DIMENSION OF   THE MODULI SPACE IN THE $\Lambda = 0$ CASE}
\par
We focus our attention on two cases~;~
a four-torus  with  flat metrics  and a
$K3$-surface  with   Calabi-Yau metrics.
\par
For a $K3$- surface with  Calabi-Yau metrics
and a four-torus with flat metrics,
the moduli space is represented only by the
deformations of a trio of  Einstein-K\"ahler forms
(hyperk\"ahler forms)
 due to our prerequisite conditions for $P$-bundle
on $M_4$  and the
gauge fixing conditions.
\par
Let $K(g)$ be the moduli space of Einstein-K\"ahler forms on $M_4$,
$\epsilon(g)$ be the moduli space of  Einstein metrics,
 and   $C(g)$ be the moduli space of  complex structures, respectively.
They are the equivalent classes  under the all diffeomorphism.
At first we   quote the result  about the dimension of $K(g)$ briefly
when $M_4$ is  a  K\"ahlerian manifold
with  vanishing  real first Chern class ,
which is    given by Ref. 23.
Then we clarify the difference between ${\cal M}(\Sigma)$, i.e. the
moduli space of hyperk\"ahler forms  defined by Eq. (\ref{eq:nineteen})
and  the moduli space ~$K(g)$.
\par
When the real first Chern class is zero,
the deformation of the K\"ahler class
with a fixed complex structure induces a deformation of a Einstein
metric from the Calabi-Yau theorem.
 The deformation of  Einstein-K\"ahler forms
$\{ \Sigma \}$  consists of
those  of Einstein metrics $\{ g \} $ and of  complex structures
$\{ J \}$ and needs a careful examination of
its degenerated part,
\begin{equation}
\delta \Sigma = \delta g \circ J + g \circ  \delta J
\sim h \circ J+ g \circ I ,
\end {equation}
where $I= {d \over dt} J(t) \mid_{t=0}$  is the  variation of complex
structure $J(t)~{\rm of}~ J$ and
$h = {d \over dt} g(t) \mid_{t=0}$  is the variation of K\"ahler-Einstein
metric $g(t)~ {\rm of}~ g$.
We  quote the results of the dimensions
about  $\epsilon(g)$, $C(g)$ and $K(g)$ in order.
\\
(1) Deformation of Einstein-K\"ahler  Metrics~:~
\par
If some infinitesimal Einstein deformations  of Einstein-K\"ahler
metric $g$  are contravariant two -tensor $h$,
then they are  decomposed into its hermitian
part $h_h$  and anti-hermitian part $h_{ah}$~;~
$\{ h \}=\{ h_h \} \oplus \{ h_{ah} \},~~$
\begin{equation}
\{ h_h \}=\{h~;~h(Ju, Jv)=h(u, v) \},~~
\{ h_{ah}\}=\{h~;~h(Ju, Jv)=-h(u, v) \},
\end{equation}
where $u,~v~ \in~ TM_4 $.
It is easy to see that both  $\{h_{h} \}$ and $\{h_{ah}\}$
 are infinitesimal Einstein deformations.
$\{ h_h \circ J \}$  are  shown to be the real (1,1) harmonic
differential 2-forms
   and orthogonal to the K\"ahler forms ( which means the
fixing of the scale factor).
Therefore they form a space  whose dimension  is
dim. $H^{(1,1)}_R(M, {\bf J})-1$.
\\
(2) Deformation of Complex Structures~:~
\par
    According to the    Kodaira-Spencer deformation theorem \cite{kodaira},
the tangent space of the moduli space  of complex structures
   is  isomorphic to 2$H^1_C(M, \Theta)$~in our case,
where  $\Theta$ is the sheaf of  the germs of
holomorphic vector fields.
The deformation  of  complex structures
 is separated into two parts.
The one is anti-symmetric complex deformation and the other is the
symmetric one.  The dimension of the anti-symmetric one
$\{ I_{as} \} $ is given by
2dim. $ H^{(2,0)}_C(M, {\bf J})$ because anti-symmetric
complex deformations are in
one to one correspondence  to (2,0) or (0,2) harmonic
differential forms of
$\{g \circ I_{as} \}$.
Thus  the dimension of the remainder $ \{I_s \} $ is given by
$2{\rm dim.}H^1_C(M, \Theta)-2{\rm dim.} H^{(2,0)}_C(M, {\bf J})$.
\\
(3) Deformation of Einstein-K\"ahler Forms~:~
\par
The degenerated part of the deformations of the Einstein-K\"ahler forms
consists of
the anti-hermitian Einstein deformations   $\{h_{ah} \circ J\}$,  and
the  symmetric complex  deformations $\{g \circ I_s\}$. The former
counterbalances  the latter   ;~~   $ g \circ I_s + h_{ah} \circ J= 0$
and  their correspondence is  shown to be bijective.
The dimension of  $\{ I_s  \} $ is the same as that of
$\{ h_{ah} \} $ and is given by
$2{\rm dim.}H^1_C(M, \Theta)-2{\rm dim.} H^{(2,0)}_C(M, {\bf J})$.
Consequently  infinitesimal deformations of the Einstein-K\"ahler  form
is represented by
\begin{eqnarray}
\delta \Sigma=
&\underbrace{
\underbrace{h_h \circ J }_{ {\rm dim.}_R H^{(1,1)}(M,{\bf J})-1 }
 + ( h_{ah} \circ J  }_ {{\rm dim.} \epsilon(g)}
\qquad \qquad \qquad & \nonumber \\
&\qquad \qquad \qquad \qquad \qquad  + \underbrace{
     g \circ I_s  )+
\underbrace {g \circ I_{as} }_{2 {\rm dim.}_C H^{(2,0)}(M,{\bf J})   }}
_{ {\rm dim.} C(g)=2 {\rm dim.}_C H^1(M, \Theta)  }&.
\end{eqnarray}
\par
Finally, the dimensions of moduli spaces  can be summarized
 for  Einstein metrics,  for complex structures and
for  Einstein-K\"ahler forms   over the
K\"ahlerian manifolds  with    $c_1(M)_R=0$,
\begin{eqnarray}
{\rm dim.} &\epsilon& (g)= {\rm dim.} H^{(1,1)}_R (M, {\bf J}) -1 +
                       2{\rm dim.}H^1_C(M, \Theta)-2{\rm dim.} H^{(2,0)}_C(M,
{\bf J}),
\\
{\rm dim.} &C& (g)= 2{\rm dim.}H^1_C(M, \Theta),
\\
{\rm dim.} &K&(g)= {\rm dim.} H^{(1,1)}_R (M,{\bf J}) -1 +
2{\rm dim.}H^1_C (M, \Theta).
\end{eqnarray}
\par
One can show that the canonical line bundle $K$ is trivial
over a $K3$-surface  or over $T^4$, and that
there is a nowhere vanishing holomorphic 2-form $ \lambda$ .
The isomorphism of sheaves due to $ \lambda$~and the Dolbeaut theorem~:
$H^1_C(M, \Theta^1) \cong H^1_C(M, \Omega^1) \cong H^{(1,1)}(M,C) $~,
leads to the following equations,
\begin{eqnarray}
{\rm dim.} &\epsilon& (g)=3 {\rm dim.} H^{(1,1)}(M,{\bf C}) -1-
2{\rm dim.} H^{(2,0)}(M,{\bf C}), \\
{\rm dim.} &C& (g)=2 {\rm dim.} H^{(1,1)}(M,{\bf C}),  \\
{\rm dim.} &K& (g)= 3{\rm dim.} H^{(1,1)}(M,{\bf C}) -1.
\end{eqnarray}
The results for a $K3$-surface and $T^4$ are given by
\[
 K3~;~ \left \{
\begin{array}{ll @{\,}}
 {\rm  dim.} &  K(g) =59,       \\
{\rm  dim.} & C(g) =40,   \\
{\rm  dim.}  & \epsilon(g)  =57,
\end{array}
\right.
\]
by substituting ~$b^{1,1}=20$ and $b^{2,0}=1$~ and~

\[
 T^4~;~ \left \{
\begin{array}{ll@{\,}}
{\rm   dim.} & K(g) = 11,   \\
{\rm  dim.} & C(g)= 8,  \\
{\rm  dim.} & \epsilon(g)  = 9,
\end{array}
\right.
\]
by substituting ~$b^{1,1}=4$ and $b^{2,0}=1$ except for a scale factor.
\par
The difference  between  ${\cal M}(\Sigma)$
and  $ K(g)$ is  as follows~;~
the moduli space of the Einstein-K\"ahler forms
$K(g)$  is defined in terms of
  $(g, J^1)$ or equivalently  $\Sigma^1$ only.
On the other hand, the definition of ${\cal M}(\Sigma)$
specifies a set of $(g, J^1, J^2, J^3)$ or  equivalently
$(\Sigma^1, \Sigma^2, \Sigma^3)$, which takes into account the
degrees of freedom  how one can choose $g$ and a trio of the $g$-orthogonal
  complex structures   up to a  scale factor.
\par
Before we  present the difference between dim. ${\cal M}( \Sigma)$
and dim. $K(g)$, let us
 show that the degrees of freedom
 of a trio of g-orthogonal complex structures
which satisfy the quaternionic relations for a fixed $g$ is 3
(see Appendix II).
\par
For a fixed $g$,
\begin{equation}
 \{ g-{\rm orthogonal}~
 {\rm quaternionic} ~{\rm almost~ complex~ structures}~J^1 \}
 \cong  S^2~  \cong ~{\rm Im} {\bf H} \mid_{x_1^2+x_2^2+x_3^2=1},
\end{equation}
where Im {\bf H} represents the imaginary part of the field of the
quaternion  {\bf H}~;
\begin{eqnarray}
{\rm Im} {\bf H}  \equiv \{J^1=x_1 \tilde J^1+x_2 \tilde J^2+ x_3
\tilde J^3 \mid &
(\tilde J^1)^2=( \tilde J^2)^2=( \tilde J^3)^2=-1,
\tilde J^1 \tilde J^2=- \tilde J^2 \tilde J^1= \tilde J^3,&
\nonumber \\ ( x_1, x_2, x_3) \in R^3
 & \tilde  J^2 \tilde J^3=- \tilde J^3 \tilde J^2= \tilde J^1,
\tilde  J^3 \tilde J^1=- \tilde J^1 \tilde J^3= \tilde J^2~
 \}.&
\end{eqnarray}
The degrees of freedom   how  one can choose $J^1$ for a fixed $g$ is given by
$  {\rm dim.}~ S^2 =2$.
The region of   $J^2 $ which is  orthogonal
to $J^1$ for a fixed pair $(g, J^1)$ is equivalent to
 $   S^1$ over $S^2$.
$J^3$ is automatically arranged after $(g, J^1, J^2)$
 are fixed.
\par
{}From  these facts,
the difference between ${\rm dim.}K(g)$ and ${\rm dim.} \epsilon(g)$
is given by 2, which corresponds to
${\rm dim.}~ S^2$ ( the degrees of freedom of how to choose $J^1$
for a given $g$) and coincides with $2{\rm dim.} H^{(2,0)}(M,{\bf C})=2$
( the degrees of freedom of how to choose
 $J^1+ \delta J^1$ for a fixed $g+ \delta g$) up to a scale factor.
The difference  between  dim. $K(g)$ and dim. ${\cal M}(\Sigma)$ is
given by dim. $S^1=1$ up to a scale factor.
\par After all the dimension of  our moduli space becomes as follows~;~
\begin{equation}
{\rm dim.}~{\cal M}(\Sigma)=60 ~{\rm for}~ K3,~~
{\rm dim.}~{\cal M}(\Sigma)=12~{\rm for}~ T^4,
\end{equation}
up to a scale factor.
In fact, we have confirmed the dimension of the moduli space
$ {\cal M}(\Sigma)$~by applying Atiyah-Singer Index theorem to
the deformation complex
( we will report this in the next paper \cite{mabe}.)
\par
The moduli space ${\cal M} (\Sigma) $ has a bundle structure
 with     the fiber $(J^1,J^2,J^3)$  over
the base manifold  which is
the moduli space of the Einstein metrics  up to a scale factor ;
\begin{eqnarray}
&{\cal M}(\Sigma)&~~~~~~~~~~
{\rm dim.}~{\cal M}(\Sigma)={\rm dim.}~K(g)+1
={\rm dim.}~ \epsilon(g)+3  \nonumber \\
&\uparrow&  \nonumber \\
& K(g)&~~~~~~~~~{\rm dim.}~ K(g)= {\rm dim.}~ \epsilon(g)+2. \nonumber \\
&\uparrow&  \nonumber \\
&\epsilon(g)& \nonumber
\end{eqnarray}
\newpage
\bf {VII. CONCLUSION }
\par
 In this paper, we have presented a topological version of 2-form
Einstein gravity in four dimensions.
 For a compact manifold in the $\Lambda \not= 0$ case,
we have defined the elliptic complex associated
with the moduli space of our theory.
 By applying the Atiyah-Singer index theorem in the $\Lambda \not= 0$
case, we have evaluated the index of the elliptic complex
and the partition function.
In the $\Lambda = 0$ case, we have  clarified  the dimension
of the moduli space
 which is related to the Plebansky's equations for $T^4$
and a $K3$-surface.
\par
 It would be  intriguing to study the $\Lambda =0$ case, since the
relation of four-dimensional (Riemann) self-dual gravity and
two-dimensional conformal field theory has been investigated.
 In fact, Park showed that the former arises from a large N-limit
of the two-dimensional sigma model with SU(N) Wess-Zumino terms
only  \cite{park}.
 Our topological model will be useful to understand the relation and
to develop the self-dual gravity.
\par
As  another approach, it would  also be  interesting to
extend BF-type model
in the  $\Lambda = 0$ case to the super BF-type model \cite{kunitomo}.
 Since  the dimension of the moduli space is non-zero,
there arise  as many fermionic zero-modes as the dimension, which
make the partition function trivial.
 To avoid this we need some functional ${\cal O}$ which absorbs the
zero-modes.
 If one calculates the vacuum expectation value of the `observable'
${\cal O}$, then it may provide non-trivial information such as a
differential invariant to distinguish differential structures on
 these  manifolds.
 Such a functional ${\cal O}$ is required to be BRST invariant to
preserve the topological nature of the theory and may be obtained from
the BRST descendant equations as in two-dimensional topological
gravity  \cite{witten3}. \par
\par
The extension of  the  algebraic curves
( one-dimensional compact complex manifolds )
with Einstein metrics   to  four dimensions
may be  the algebraic surfaces
(two-dimensional compact  complex   manifolds )
with Einstein metrics.
$T^4$ and a $K3$-surface   belong to   the  algebraic surfaces.
To construct the topological gravity models
by taking another gauge fixing
conditions, which describe these
algebraic surfaces is worth pursuing.
\vskip 2.0cm
\newpage
\begin{center}
{\bf ACKNOWLEDGMENT}
\end{center}
 \ We are grateful to  Q-Han Park and S. Morita
for useful discussions.
 We also acknowledge  N. Sakai for useful discussions and
careful reading of  the manuscript.
One of us (M. A.)  thanks  A. Futaki  most for pointing out
   the difference between $K(g)$  and ${\rm M} (\Sigma)$ and the
necessity of  an  almost complex structure with  vanishing real
first Chern class for the $\Lambda =0$ case.
\newpage
\appendix{APPENDIX I~:~ PROPERTIES OF THE $\eta$ SYMBOLS}
\par
We list some useful identities of the $\eta^k_{ab}$  \cite{t'hooft}:~
\begin{equation}
\eta^k_{ab}=\epsilon_{kab}~{\rm for} ~a,~b=1,2,3.~~~
\eta^k_{a0}=\eta^k_{0a}=\delta^k_a~{\rm for} ~a=0,1,2,3.
\nonumber
\end{equation}
\begin{eqnarray}
&(1)& \eta^k_{ab}=-\eta^k_{ba},
\nonumber \\
&(2)& \eta^k_{ab}=-{1 \over 2}\epsilon_{ab}{}^{cd}\eta^k_{cd},
\nonumber \\
&(3)& \eta^k_{ab}\eta^k_{cd}
=2(\delta_a^{[c} \delta_b^{d]} -{1 \over 2}\epsilon_{abcd})
\equiv 2P_{ab}{}^{cd},
\nonumber \\
&(4)& \eta^k_{ab}\eta_c^{lb}
=2\delta_k^l \delta_c^a +\epsilon_{kln}\eta^n_{ac},
\end{eqnarray}
where $\epsilon_{klm}$ and $\epsilon_{abcd}$
 denote the  anti-symmetric constant tensors.
\par
{}From  Eqs.(3) and (4), the  identities of $\Sigma^k(e)$ are
derived~:~
\begin{eqnarray}
&(5)& \Sigma^k_{\mu \nu}(e)\Sigma^k{}^{\tau \rho}(e)
=2P_{\mu \nu}{}^{\rho \tau},
\nonumber \\
&(6)& \Sigma^k_{\mu}{}^{\nu}(e) \Sigma^l_{\rho \nu}(e)
=-\delta_k^l \delta_\mu^\rho +\epsilon_{kln}\Sigma^n_{\mu}{}^{\rho}(e).
\nonumber
\end{eqnarray}
\newpage
\appendix{APPENDIX II ~:~ SOME DEFINITIONS, THEOREMS AND PROPOSITION}
\par
In this appendix, we put some definitions, theorems and proposition
which we have  used.
\\
\\
Theorem ( S. Kobayashi  and K. Nomizu \cite{kobayashi} )
\\
Let $P(M, G)$  be a principal fibre bundle with a connection $\Gamma$,
where $M$ is connected and paracompact. Let $u_0$ be an arbitrary
point  $P$. Denote by $P(u_0)$ the set of points in $P$ which
can be joined to $u_0$ by a horizontal curve. Then
\\
(1) $P(u_0)$ is a reduced bundle with structure group $Hol(g)$.\\
(2) The connection $\Gamma$ is reducible to a connection in $P(u_0)$.
\\
\\
Theorem ( E. Calabi and Yau \cite{calabi} )
\\
Let $M$ be a compact K\"ahlerian manifold, $\Sigma$ is
its K\"ahler form  and
Any closed (real)
2-form of type (1,1) belonging to $2\pi c_1(M)_R$ is the Ricci form
of one and only one K\"ahler class $\Sigma$.
\par
As an immediate consequence, we get the following fact:
\\
the compact K\"ahlerian manifolds with zero real first Chern
class are exactly the compact complex manifolds admitting a K\"ahler
metric with zero Ricci form (equivalently with restricted holonomy group
contained in the special unitary group.)
\\
\\
Definition 1 (Besse \cite{besse})
\\
A $4n$-dimensional Riemannian manifold is called
\\
(a) hyperk\"ahlerian if its holonomy group is contained in $SP(n)$.
\\
(b) locally hyperk\"ahlerian if its restricted holonomy is contained in
$SP(n)$.
\\
\\
Definition  2 (Besse \cite{besse})
\\
A $4n$-dimensional Riemannian manifold is called
\\
(a) quaternion-K\"ahler if its holonomy group is contained in
 $SP(N) \times SP(1)$
\\
(a)locally  quaternion-K\"ahler if its restricted holonomy
group is contained in  $SP(N) \times SP(1)$
\\
\\
Proposition 1 (Besse \cite{besse})
\\
A Riemannian manifold $ (M, g) $ is hyperk\"ahlerian if and only
if there exist on $M$  two complex structures $J^1$ and $J^2$
 such that
\\
(a) $J^1$ and $J^2$ are parallel (i.e. $g$ is a K\"ahler metric for each. )
\\
(b) $J^1 J^2= -J^2 J^1$
\par
Notice that $J^3$ is still a parallel complex structure on $M$
and more generally, given $(x_1,x_2,x_3)$ in $R^3$ with
$ x_1^2+x_2^2+x_3^2=1$,
then the complex structure $J=x_1J^1+x_2J^2+x_3J^3$ on $M$ is still
parallel. So there is a whole manifold  (isomorphic to $S^2$ ) on parallel
complex structure on $M$.
\\
\\
\par
\newpage



\begin{center}
\begin{tabular}{|lr|| lr|}  \hline \hline
\multicolumn{4}{|c|}{Table 1. A Dimension-Counting
of Fundamental Variables. } \\ \hline
\it $\Lambda \not = 0$~{\rm  case}   & degrees
&$\Lambda  = 0$~case    &  degrees    \\ \hline\hline
\it $ \Sigma^k $ & $3 \times 6= 18 $
       & $ \Sigma^k $& $3 \times 6= 18$  \\ \hline
\it $ \omega^k $ & $ 3 \times 4= 12 $
 & $ \omega^k$ & $1 \times 4= 4 $ \\ \hline
\it {\rm total} & 30   & total & 22 \\ \hline\hline
\it \{ diffeo. $ \times SU(2)$ \}  & {}
   & \{ diffeo. $ \times U(1)$ \}  & {}
\\
\it {\rm gauge~fix.~condi.}  &
4+3=7    & gauge~fix.~condi.  &4+1=5
\\ \hline
\it \{ {\rm rest.~ top. / red.} \}  & {}
    & \{  rest.~ top. / red. \}  & {}
 \\
\it  {\rm gauge~ fix.~ condi. } & 5
    &  gauge~ fix.~ condi. &~ 5
 \\ \hline
\it {\rm Eq. of the  motion}  &{}
 & Eqs. of  motion & {} \\
\it $ F^k=( \Lambda/12) \Sigma^k$ & 18
    & $\{F^k=0\}/\{DF^k=0\}/$ &
 \\
\it  & {}
    & $\{D^2F^k=0\}$, & 6-4+1=3
 \\
\it   & {} &
$\{D \Sigma^k=0\}/\{D^2\Sigma^k=0\}$& $ 3 \times (4-1)=9$
     \\ \hline
\it {\rm total}  & 30
 & total  & 22 \\ \hline\hline
\end{tabular}
\end{center}
\small{
\begin{center}
\begin{tabular}{|l|l|l|l| l|r|} \hline\hline
\multicolumn{6}{|c|}{Table 2.~~ Fields and Their Ghost Assignment}
\\ \hline
\it field & content  &  Fermion/  & ghost  & form  & represen-\\
\it       &          &  Boson        & number   &       &tation       \\
\hline\hline
\it $ \delta\omega^i=\delta\omega^i_{\mu\nu}dx^\mu\wedge dx^\nu$ &{} &
B  & 0 & 1 & $\Omega^{2, 0} \otimes \Lambda^1$ \\ \hline
\it $\delta\Sigma^i=\delta  \Sigma_{\mu\nu}dx^\mu \wedge dx^\nu $ &{} &
 B& 0 & 2 & $\Omega^{2, 0} \otimes \Lambda^2$ \\ \hline\hline
\multicolumn{6}{|l|}{ diffeo. $\rightarrow$ BRST } \\ \hline
\it $ c^\nu$&  ghost  &
F    & 1 &-1 & $ TM_4  \simeq \Lambda^1$ \\ \hline
\it $b=b_\nu dx^\nu$&  anti-ghost    &
F    &-1 & 1 &$ T^\ast M_4  \simeq \Lambda^1$ \\ \hline
\it $\pi= \pi_\nu dx^\nu$& N-L~field  &
B &0 & 1 &$ T^\ast M_4  \simeq \Lambda^1$ \\ \hline\hline
\multicolumn{6}{|l|}{SU(2) $\rightarrow$ BRST} \\ \hline
\it $c^i$~& ghost &
F    & 1 & 0 &$ \Omega^{2, 0} \otimes \Lambda^0$ \\ \hline
\it $b^i$&anti-ghost  &
F    &-1 & 0 & $\Omega^{2, 0} \otimes \Lambda^0$ \\ \hline
\it $\pi^i$& N-L~field &
B & 0 & 0 & $\Omega^{2, 0} \otimes \Lambda^0$ \\ \hline\hline
\multicolumn{6}{|l|}{red.~ diffeo. $\rightarrow$ BRST } \\ \hline
\it $\gamma^\nu$ & ghost  &
B    & 2 & -1&  $TM_4  \simeq \Lambda^1$ \\ \hline
\it $\beta= \beta_\nu dx^\nu$& anti-ghost &
B &-2 & 1 & $T^\ast M_4  \simeq \Lambda^1$ \\ \hline
\it $ \tau=\tau_\nu dx^\nu$&N-L~field &
F    &-1 & 1 & $T^\ast M_4  \simeq \Lambda^1$ \\ \hline\hline
\multicolumn{6}{|l|}{ red.~SU(2) $\rightarrow$ BRST} \\ \hline
\it $\gamma^i$&ghost  &
B    & 2 & 0 & $ \Omega^{2, 0} \otimes \Lambda^0$ \\ \hline
\it $\beta^i$~& anti-ghost  &
B &-2 & 0 &$ \Omega^{2, 0} \otimes \Lambda^0$ \\ \hline
\it $\tau^i$&  N-L~field &
F    & -1 & 0 & $\Omega^{2, 0} \otimes \Lambda^0$ \\ \hline\hline
\multicolumn{6}{|l|}{susy. $\rightarrow$ BRST} \\ \hline
\it $\phi^i=\phi^i_\nu dx^\nu$&ghost &
F    & 1 & 1 &$ \Omega^{2, 0} \otimes  \Lambda^1$ \\ \hline
\it  $\chi^{ij}$& anti-ghost  &
F &-1 & 0 &$ \Omega^{4,0} \times \Lambda^0$ \\ \hline
\it $ \pi^{ij}$& N-L~ field  &
B    & 0 & 0 & $ \Omega^{4,0} \times \Lambda^0$ \\ \hline\hline
\end{tabular}
\end{center}
}
{}~~~~~~~~~~~~~~~~~~~~~~~~~~~~~~~~~~~~~~~~~~~~~~~~~~~~
{}~~~~~~~~~( $-1$ for form means a "vector")

\newpage
Figure caption

Fig. 1~~~  $Hol(g)$~  on $M_4$ with torsionless connections

1. almost complex~ $\{ M_4, J\}$~

2. K\"ahlerian~ $\{M_4, g, J\}$~

3.  Riemannian~ $\{M_4, g \}$~
 ~$O(4) \supseteq Hol(g)$
 ~e. g. $S^4$ with Riemannian metrics

4. Ricci-flat K\"ahlerian~$\{M_4, g, J\}$~

5. hyperk\"ahlerian~$\{M_4, g, J^1, J^2, J^3 \} $~
\newpage
\begin{center}
\bf{Fig. 1~}
\end{center}
\begin{picture}(200,300)
\put(150,200){\circle{40}}
\put(175,200){\circle{40}}
\put(163,200){\circle{13}}
\put(161,200){\circle{6}}
\put(135,200){1}
\put(175,200){3}
\put(152, 160){\vector(1,3){10}}
\put(150,148){2}
\put(135, 178){\vector(3,2){28}}
\put(130,165){4}
\put(126, 192){\vector(4,1){35}}
\put(115,182){5}
\end{picture}
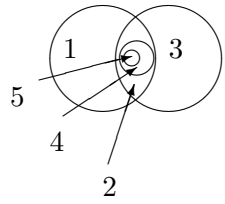
\end{document}